\documentclass[aps,prl,twocolumn,showpacs,preprintnumbers,amsmath,amssymb,superscriptaddress]{revtex4}%

\usepackage{graphicx}%
\usepackage{dcolumn}
\usepackage{amsmath}
\usepackage{color}

\begin{document}

\preprint{PREPRINT (\today)}

\title{Muon-spin rotation study of the in-plane magnetic penetration depth of FeSe$_{0.85}$: evidence for nodeless superconductivity}

\author{R.~Khasanov}
 \email{rustem.khasanov@psi.ch}
 \affiliation{Laboratory for Muon Spin Spectroscopy, Paul Scherrer
Institut, CH-5232 Villigen PSI, Switzerland}
\author{K.~Conder}
 \affiliation{Laboratory for Developments and Methods, Paul Scherrer Institute,
CH-5232 Villigen PSI, Switzerland}
\author{E.~Pomjakushina}
 \affiliation{Laboratory for Developments and Methods, Paul Scherrer Institute,
CH-5232 Villigen PSI, Switzerland}
 \affiliation{Laboratory for Neutron Scattering, Paul Scherrer Institute \& ETH
 Zurich,CH-5232 Villigen PSI, Switzerland}
\author{A.~Amato}
 \affiliation{Laboratory for Muon Spin Spectroscopy, Paul Scherrer
Institut, CH-5232 Villigen PSI, Switzerland}
\author{C.~Baines}
\affiliation{Laboratory for Muon Spin Spectroscopy, Paul Scherrer
Institut, CH-5232 Villigen PSI, Switzerland}
\author{Z.~Bukowski}
 \affiliation{Laboratory for Solid State Physics, ETH Z\"urich, CH-8093 Z\"urich,
Switzerland}
\author{J.~Karpinski}
 \affiliation{Laboratory for Solid State Physics, ETH Z\"urich, CH-8093 Z\"urich,
Switzerland}
\author{S.~Katrych}
 \affiliation{Laboratory for Solid State Physics, ETH Z\"urich, CH-8093 Z\"urich,
Switzerland}
\author{H.-H.~Klauss}
 \affiliation{IFP, TU Dresden, D-01069 Dresden, Germany}
\author{H.~Luetkens}
 \affiliation{Laboratory for Muon Spin Spectroscopy, Paul Scherrer
Institut, CH-5232 Villigen PSI, Switzerland}
\author{A.~Shengelaya}
 \affiliation{Physics Institute of Tbilisi State University,
Chavchavadze 3, GE-0128 Tbilisi, Georgia}
\author{N.D.~Zhigadlo}
 \affiliation{Laboratory for Solid State Physics, ETH Z\"urich, CH-8093 Z\"urich,
Switzerland}

\begin{abstract}
The in-plane magnetic penetration depth $\lambda_{ab}$ of the iron
selenide superconductor with the nominal composition FeSe$_{0.85}$
was studied by means of muon-spin rotation. The measurements of
$\lambda_{ab}^{-2}(T)$ are inconsistent with the presence of nodes
in the gap as well as with a simple isotropic $s-$wave type of the
order parameter, but can be equally well described within a
two-gap ($s+s$) and an anisotropic $s-$wave gap picture. This
implies that the superconducting energy gap in FeSe$_{0.85}$ contains no nodes.
\end{abstract}
\maketitle

The recent discovery of the Fe-based high-temperature
superconductors has attracted considerable attention to the
pnictides. The superconductivity was first found in 
LaO$_{1-x}$F$_x$FeAs \cite{Kamihara08} and, later on, in other
single-layer arseno-pnictides by replacing La with various rare-earth elements
(Sm, Nd, Pr, Ce, Ho, Y, Dy, and Tb)
\cite{Wang08_Ren08_Sefat08_Chen08_Wen08_Yang08} as well as in
oxygen-free compounds such as doped double-layer MFe$_2$As$_2$ (M=Ba, Sr, and Ca)
\cite{Rotter08_Chen08_Ni_Torikachvili08} and single-layer LiFeAs
\cite{Wu_Pitcher_Tapp08}. The common structural feature of all these
materials is the Fe-As layers consisting of an Fe square planar
sheet tetrahedrally coordinated by As. Recently, superconductivity
with the transition temperature $T_c\simeq8$~K was discovered in
$\alpha-$FeSe with PbO-structure \cite{Hsu08}. This compound also
has a Fe square lattice with Fe atoms tetrahedrally coordinated by
Se ones, similar to the structure of FeAs planes in the single- and
the double-layer arseno-pnictides. In this respect FeSe, consisting
of the ''superconducting`` Fe-Se layers only, can be treated as a
prototype of the known families of Fe-As based high-temperature
superconductors and, consequently, becomes a good modeling system
to study the mechanism leading to the occurrence of
superconductivity in these new class of materials.

Here we report a study of the in-plane magnetic field penetration
depth $\lambda_{ab}$ in iron selenide superconductor with the
nominal composition FeSe$_{0.85}$ by means of muon-spin rotation
($\mu$SR). $\lambda_{ab}^{-2}(T)$ was
reconstructed from the temperature dependences of the $\mu$SR
linewidth measured in a magnetic field of 0.01~T. The observed
$\lambda_{ab}^{-2}(T)$ was found to be equally well described within
the framework of anisotropic $s-$wave as well two-gap $s+s-$wave
models. In a case of anisotropic $s-$wave model the maximum value of
the gap at $T=0$ was found to be $\Delta_0=1.35$~meV leading to
$2\Delta_0/k_BT_c=3.79$, close to the weak-coupling BCS value 3.52.
The two-gap $s+s-$wave model yields $\Delta_{0,1}=1.63$~meV and
$\Delta_{0,2}=0.38$~meV. The corresponding gap to $T_c$ ratios are
$2\Delta_{0,1}/k_BT_c=4.59$ and $2\Delta_{0,2}/k_BT_c=1.07$ close to
those reported for various single- and double-layer arseno-pnictide
superconductors \cite{Szabo08,Gonnelli08,Ren08_2}.

\begin{figure}[htb]
\includegraphics[width=0.8\linewidth]{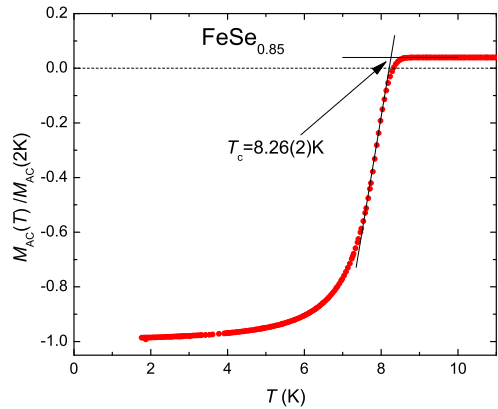}
\caption{(Color online) Temperature dependence of the AC
magnetization $M_{AC}$ ($\mu_0H_{AC}=0.1$~mT, $\nu=1000$~Hz) of
FeSe$_{0.85}$. }
 \label{fig:magnetization}
\end{figure}

Details on the sample preparation for FeSe$_{0.85}$ can be  found
elsewhere \cite{Hsu08}. X-ray diffraction analysis reveals that the
$\alpha$-FeSe phase is dominant and that the amount of the impurity
fraction does not exceed $\simeq$7-10\%.
The AC magnetization ($M_{AC}$) measurements ($\mu_0H_{AC}=0.1$~mT,
$\nu=1000$~Hz) were performed on a Quantum Design PPMS magnetometer at
temperatures ranging from $1.75$~K to 300~K. The superconducting
transition temperature $T_c=8.26(2)$~K was obtained as an
intersection of the linearly extrapolated $M_{AC}(T)$ with $M_{AC} =
const$  line (see Fig.~\ref{fig:magnetization}). 

Zero field (ZF), longitudinal field (LF) and transverse field (TF)
$\mu$SR  experiments were performed at the
$\pi$M3 beam  line at the Paul Scherrer Institute (Villigen,
Switzerland). Here LF and TF denote the cases when the magnetic
field is applied parallel and perpendicular to the initial muon-spin
polarization, respectively. The experiments down to $T\simeq1.5$~K
were performed on the low-background General Purpose Surface-Muon
instrument (GPS) and those down to $T\simeq0.02$~K on the
Low-Temperature-Facility (LTF) instrument.

\begin{figure}[htb]
\includegraphics[width=0.8\linewidth]{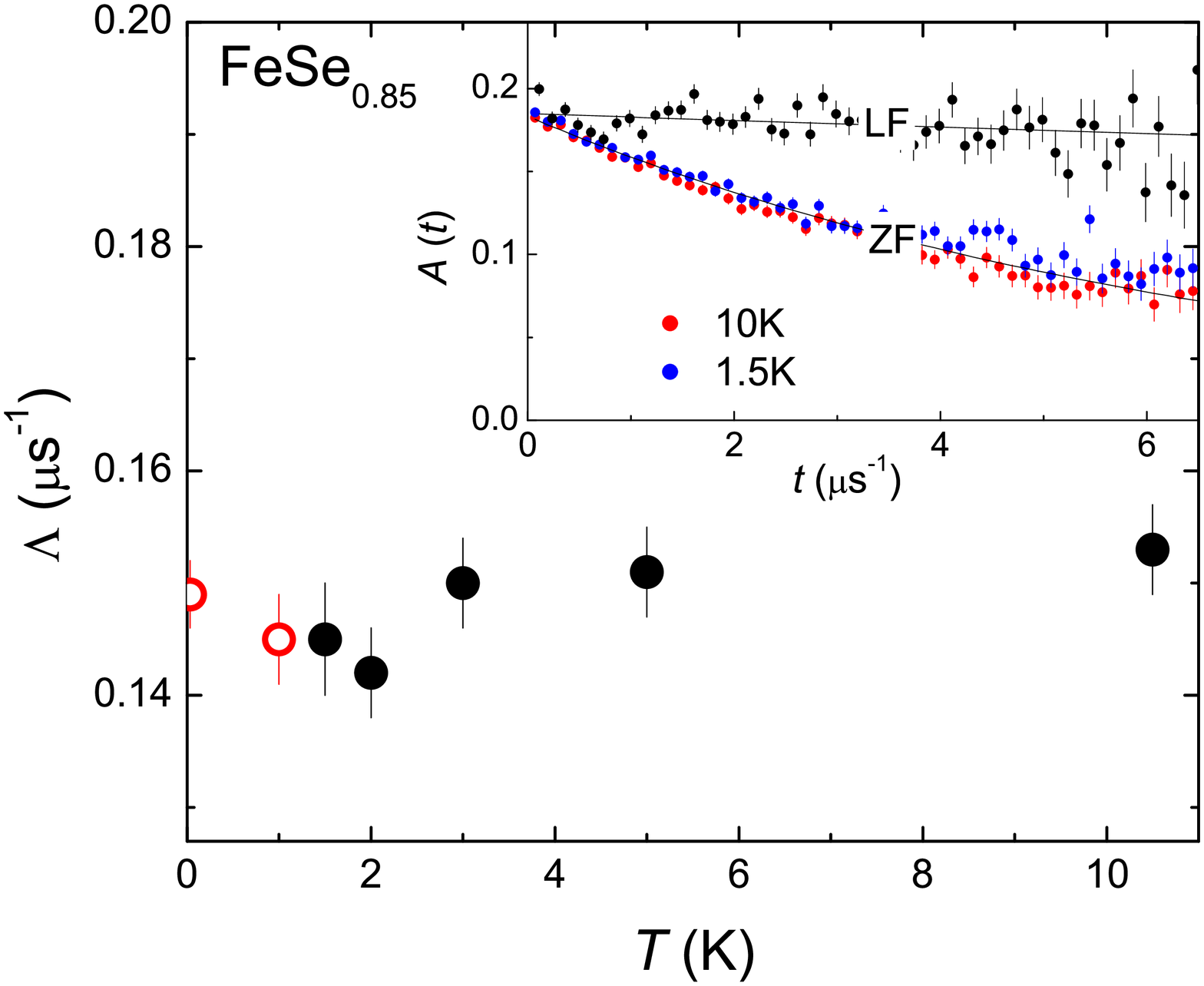}
\caption{(Color online) Temperature dependence of the ZF muon
depolarization rate $\Lambda$ of FeSe$_{0.85}$. The inset shows ZF
($T=1.5$~K and 10~K) and LF ($T=10~K$, $\mu_0H=0.01$~T) $\mu$SR
time-spectra of FeSe$_{0.85}$.}
 \label{fig:ZF-LF}
\end{figure}

First we are going to present the results of the ZF experiments. In
the whole temperature region the ZF data were found to be well
described by the single-exponential decay function:
\begin{equation}
A^{ZF}(t)=A_0 \exp(-\Lambda t).
 \label{eq:ZF-LF}
\end{equation}
Here $A_0$ is the initial asymmetry at $t=0$ and $\Lambda$  is the
exponentional depolarization rate. The results of the analysis of
the ZF data and the representative ZF and LF muon-time spectra are
shown in Fig.~\ref{fig:ZF-LF}. The open and the closed symbols are
from the measurements taken on the LTF and the GPS instruments,
respectively.

The exponential character of the muon polarization decay might be
explained either by existence of fast electronic fluctuations
measurable within the $\mu$SR time-window \cite{Khasanov08_Sm-Nd} or
by a static magnetic field distribution caused by diluted and randomly
oriented magnetic moments \cite{Walstedt74}. To distinguish between
these two cases LF $\mu$SR experiments were performed. As is shown
in Ref.~\onlinecite{Schenck86} in a case when the applied
longitudinal field is much stronger than the internal field
($B>10B_{int}$) the muon spins become ''decoupled`` from the static
internal field. On the other hand, field fluctuations perpendicular
to the applied external field can cause irreversible spin-flip
transitions of the muon spin, leading to depolarization
\cite{Khasanov08_Sm-Nd}. As is shown in the inset of
Fig.~\ref{fig:ZF-LF} an external field of 0.01~T is
already enough to completely decouple the muon spins. This proves
that the magnetism in FeSe$_{0.85}$ sample studied here is static in
origin and is caused by diluted and randomly distributed magnetic moments.
Bearing this in mind and by taking into account the presence of the
relatively high paramagnetic contribution at $T>T_c$ (see
Fig.~\ref{fig:magnetization}) we may conclude that the magnetism
observed in both, ZF $\mu$SR and magnetization experiments, has
similar sources and, most probably, caused by the traces of Fe
impurities \cite{Hsu08}.

\begin{figure}[htb]
\includegraphics[width=0.8\linewidth]{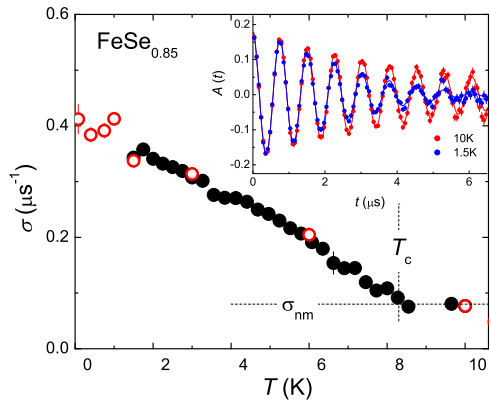}
\caption{(Color online) Temperature dependence of  the Gaussian
depolarization rate $\sigma$ at $\mu_0H=0.01$~T of FeSe$_{0.85}$. The inset shows the
TF muon-time spectra above ($T=10$~K) and below ($T=1.5~K$) the
superconduting transition temperature $T_c=8.26$~K. }
 \label{fig:TF}
\end{figure}

\begin{figure*}[htb]
\includegraphics[width=0.65\linewidth]{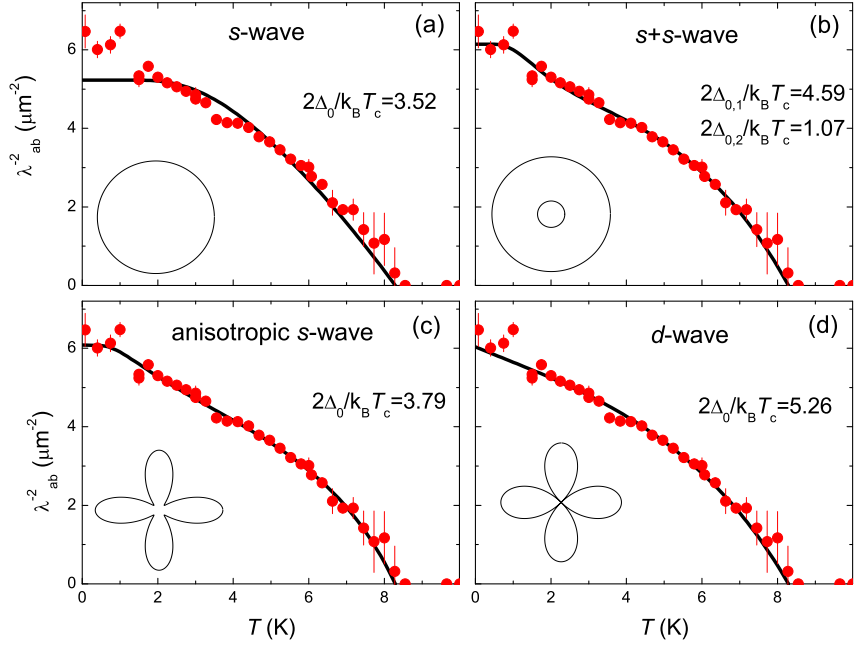}
\caption{(Color online) Temperature dependence  of
$\lambda_{ab}^{-2}$ of FeSe$_{0.85}$ obtained from measured
$\sigma_{sc}$  by means of Eq.~(\ref{eq:lambda_ab}). The fitting
curves (solid black lines) were obtained within the following
models of the gap symmetries: $s-$wave (a), $s+s-$wave (b),
anisotropic $s-$wave (c), and $d-$wave (d). The corresponding
angular dependences of the gaps are shown as the insets. }
 \label{fig:gap_analysis}
\end{figure*}

The in-plane magnetic penetration depth $\lambda_{ab}$ was  studied
in the TF $\mu$SR experiments. In a powder sample the magnetic
penetration depth $\lambda$ can be extracted from the Gaussian
muon-spin depolarization rate $\sigma_{sc}(T) \propto
1/\lambda^{2}(T)$, which probes the second moment of the magnetic
field distribution in the mixed state \cite{Zimmermann95}. For
highly anisotropic layered superconductors (like the pnictide
superconductors) $\lambda$ is mainly determinated by the in-plane
penetration depth $\lambda_{ab}$ \cite{Fesenko91}: $ \sigma_{sc}(T)
\propto 1/\lambda_{ab}^{2}(T)$. By taking into account the weak
magnetism observed in our ZF experiments (see Fig.~\ref{fig:ZF-LF}
and discussion above) the TF $\mu$SR data were analyzed by using the
following functional form:
\begin{equation}
A^{TF}(t)=A_0\exp(-\Lambda t)\exp(-\sigma^2t^2/2)\cos(\gamma B_{int}t+\phi).
 \label{eq:TF}
\end{equation}
Here $\gamma/2\pi= 135.5$~MHz/T is the muon gyromagnetic ratio,
$\phi$ is the initial phase of the muon-spin ensemble, and
$\sigma=(\sigma_{sc}^2+\sigma_{nm}^2)^{0.5}$ is the Gaussian
relaxation rate. $\sigma_{nm}$ is the nuclear magnetic dipolar
contribution which is generally temperature independent
\cite{Schilling82}. The whole set of 0.01~T TF $\mu$SR data was fitted
simultaneously with $A_0$, $\Lambda$, and $\phi$ as a common
parameters and $\sigma$ and $B_{int}$ as individual parameters for
each temperature point. The exponential relaxation rate was
assumed to be temperature independent in accordance with the
results of our ZF $\mu$SR experiments (see Fig.~\ref{fig:ZF-LF}).
$\sigma_{nm}$ was fixed to the value obtained above $T_c$ where
$\sigma=\sigma_{nm}$ (see Fig.~\ref{fig:TF}). The results of the
analysis and the representative TF muon-time spectra are shown in
Fig.~\ref{fig:TF}. The open and the closed symbols are again from
the measurements taken on the LTF and the GPS instruments,
respectively.

The superconducting part of the Gaussian depolarization rate
$\sigma_{sc}$ can be converted into $\lambda_{ab}$ via
\cite{Khasanov08_Sm-Nd,Brandt88,Fesenko91}:
\begin{equation}
\sigma_{sc}^2/\gamma_\mu^2=0.00126\Phi_0^2/\lambda_{ab}^{\ 4},
 \label{eq:lambda_ab}
\end{equation}
where $\Phi_0=2.068\cdot10^{-15}$~Wb is the magnetic flux quantum.
Fig.~\ref{fig:gap_analysis} shows $\lambda^{-2}_{ab}(T)$ obtained
from the measured $\sigma_{sc}(T)$ by means of
Eq.~(\ref{eq:lambda_ab}). Regarding the pairing symmetry, available
experimental results on various single- and double-layer
arseno-pnictides are divided between those favoring an isotropic
\cite{Ding08,Hashimoto08} as well as an anisotropic
\cite{Kondo08_Martin08} nodeless gap and those supporting line nodes
\cite{Boyer08_Ren08_Shan08_Grafe08_Nakai08_Mukuda08}. The two-gap
behavior was also reported in
Refs.~\onlinecite{Ren08_2,Ding08,Szabo08,Malone08,Evtushinsky08}.
Bearing this in mind the data in Fig.~\ref{fig:gap_analysis} were
analyzed by using single-gap and two-gap models, assuming that the
superconducting energy gaps have the following symmetries: $s-$wave
(a), $s+s-$wave (b), anisotropic $s-$wave (c), and $d-$wave (d).

Temperature dependence of the magnetic
penetration depth $\lambda$ was calculated within the local
(London) approximation ($\lambda\gg\xi$) by using the following
functional form \cite{Tinkham75,Khasanov07_La214}:
\begin{equation}
\frac{\lambda^{-2}(T)}{\lambda^{-2}(0)}=  1
+\frac{1}{\pi}\int_{0}^{2\pi}\int_{\Delta(T,\varphi)}^{\infty}\left(\frac{\partial
f}{\partial E}\right)\frac{E\
dEd\varphi}{\sqrt{E^2-\Delta(T,\varphi)^2}}.
 \label{eq:lambda-d}
\end{equation}
Here $\lambda^{-2}(0)$ is the zero-temperature value  of the
magnetic penetration depth, $f=[1+\exp(E/k_BT)]^{-1}$ is  the
Fermi function, $\varphi$ is the angle along the Fermi surface,
and $\Delta(T,\varphi)=\Delta_0 \delta(T/T_c)g(\varphi)$
($\Delta_0$ is the maximum gap value at $T=0$).  The temperature
dependence of the gap is approximated by
$\delta(T/T_c)=\tanh\{1.82[1.018(T_c/T-1)]^{0.51}\}$
\cite{Carrington03}. The function $g(\varphi)$ describes the
angular dependence of the gap and is given by $g^s(\varphi)=1$ for
the $s-$wave gap, $g^d(\varphi)=|\cos(2\varphi)|$ for the
$d-$wave gap, and $g^{s_{An}}(\varphi)=(1+a\cos4\varphi)/(1+a)$
for the anisotropic $s-$wave gap \cite{Shan05}.

The two-gap calculations were performed within the  framework of
the so called $\alpha-$model assuming that the total superfluid
density is a sum of two components
\cite{Khasanov07_La214,Carrington03}:
\begin{equation}
\frac{\lambda^{-2}(T)}{\lambda^{-2}(0)}=
\omega\cdot\frac{\lambda^{-2}(T,
\Delta_{0,1})}{\lambda^{-2}(0,\Delta_{0,1})}+(1-\omega)\cdot
\frac{\lambda^{-2}(T, \Delta_{0,2})}{\lambda^{-2}(0,\Delta_{0,2})}.
\end{equation}
Here $\Delta_{0,1}$ and $\Delta_{0,2}$ are  the zero-temperature
values of the large and the small gap, respectively, and $\omega$
($0\leq\omega\leq1$) is the weighting factor which represents the
relative contribution of the larger gap to  $\lambda^{-2}$.

The results of the analysis are presented  in
Fig.~\ref{fig:gap_analysis} by solid black lines.  The angular
dependences of the gaps [$\Delta_0\cdot g(\varphi)$] are shown in
the corresponding insets.  It is obvious that the simple $s-$ and
$d-$wave approaches cannot describe the observed
$\lambda^{-2}_{ab}(T)$ [see Figs.~\ref{fig:gap_analysis}~(a) and
(d)]. In both cases the low temperature points stay systematically
higher than the theoretically derived curves. The constant, within
our experimental uncertainty, $\lambda_{ab}^{-2}(T)$ at
$T\lesssim1$~K is also inconsistent with the presence of any type of
nodes in the energy gap of FeSe$_{0.85}$.    In contrast, both,
anisotropic $s-$ and two-gap $s+s-$wave models
[Figs.~\ref{fig:gap_analysis}~(b), (c)] describe the experimental
data reasonably well. In the following we are going to discuss
separately the results obtained within the framework of these two
models. For the anisotropic $s-$wave case we get
$\Delta_0=1.35$~meV, $a=0.796$, and $\lambda_{ab}(0)=406$~nm. The
corresponding gap to $T_c$ ratio is $2\Delta/k_BT_c=3.79$ which is
rather close to the weak-coupling BCS value 3.52. The obtained
variation with angle $\Delta^{max}/\Delta^{min}=
1.35/0.153\simeq8.8$ is substantially bigger than 1.2 reported in
Ref.~\onlinecite{Kondo08_Martin08} for NdFeAsO$_{0.9}$F$_{0.1}$.
The fit within the two-gap $s+s-$wave model yields
$\Delta_{0,1}=1.63$~meV, $\Delta_{0,2}=0.38$~meV, $\omega=0.654$,
and $\lambda_{ab}(0)=403$~nm. It is interesting to note that the
''large`` and the ''small`` gap to $T_c$ ratios
$2\Delta_{0,1}/k_BT_c=4.59$ and $2\Delta_{0,2}/k_BT_c=1.07$ are
very close to those reported for various single- and double-layer arseno-pnictide superconductors based on the results of the
point contact Andreev reflection spectroscopy experiments of Szabo
{\it et al.} \cite{Szabo08} and Gonelli {\it et al.}
\cite{Gonnelli08}, and the first critical field measurements of Ren
{\it et al.} \cite{Ren08_2}. The multiple gaps may originate from
the multiple bands at the Fermi level of FeSe$_{0.85}$. First-principle calculation indicates that the
Fermi surface (FS) of FeSe is quasi-two dimensional and consists of
hole-type sheets around the $\Gamma$ point and electron-type sheets
around the M point of the Brillouin zone \cite{Subedi08}. It is
conceivable that the two gaps open up on the different sheets of the
FS. In this context, the present compound may resemble the situation
of MgB$_2$ where a large gap opens on the FS derived from the
orbitals in the boron plane, while a small gap opens on the FS
derived from orbitals perpendicular to the boron plane
\cite{Choi02}.

To conclude, muon-spin rotation measurements  were performed on
the superconductor  FeSe$_{0.85}$ ($T_c\simeq8.3$~K).
$\lambda_{ab}^{-2}(T)$ was reconstructed from the temperature
dependence of the Gaussian muon depolarization rate measured at
$\mu_0H=0.01$~T. The absolute value of the in-plane magnetic
penetration depth $\lambda_{ab}$ at $T=0$ was estimated to be
$\lambda_{ab}(0)\simeq405$~nm. The temperature dependence of
$\lambda_{ab}^{-2}$  was found to be inconsistent with an
isotropic $s-$wave as well as with a $d-$wave symmetry of the
superconducting energy gap. A good agreement between the
experimental data and the theory was obtained within the framework
of a two-gap $s+s$ and an anisotropic $s-$wave gap models thus
suggesting that the superconducting energy gap in FeSe$_{0.85}$
superconductor is fully developed and contains no nodes.

This work was performed at the Swiss Muon Source (S$\mu$S),  Paul
Scherrer Institute (PSI, Switzerland). The work was partly supported by the NCCR program
MaNEP.

\end{document}